\documentclass{article}
\usepackage[preprint]{neurips_2022}
\usepackage{amsmath,amsfonts}
\usepackage[ruled]{algorithm2e}
\usepackage{array}
\usepackage[caption=false,font=normalsize,labelfont=sf,textfont=sf]{subfig}
\usepackage{textcomp}
\usepackage{stfloats}
\usepackage{url}
\usepackage{verbatim}
\usepackage{graphicx}
\usepackage{bm}
\usepackage{booktabs}
\usepackage{multirow}
\usepackage{dsfont}

\begin{document}
	
\title{Patch-Based Denoising Diffusion Probabilistic Model for Sparse-View CT Reconstruction}

\author{Wenjun Xia, Wenxiang Cong, Ge Wang\\
	\normalfont Department of Biomedical Engineering \\
	Rensselaer Polytechnic Institute\\
	Troy, NY 12180 USA\\
	\texttt{xiaw4@rpi.edu, wxcong@gmail.com, wangg6@rpi.edu}
}



\maketitle

\begin{abstract}
	Sparse-view computed tomography (CT) can be used to reduce radiation dose greatly but is suffers from severe image artifacts. Recently, the deep learning-based method for sparse-view CT reconstruction has attracted a major attention. However, neural networks often have a limited ability to remove the artifacts when they only work in the image domain. Deep learning-based sinogram processing can achieve a better anti-artifact performance, but it inevitably requires feature maps of the whole image in a video memory, which makes handling large-scale or three-dimensional (3D) images rather challenging. 
	In this paper, we propose a patch-based denoising diffusion probabilistic model (DDPM) for sparse-view CT reconstruction.
	A DDPM network based on patches extracted from fully sampled projection data is trained and then used to inpaint down-sampled projection data. The network does not require paired full-sampled and down-sampled data, enabling unsupervised learning. Since the data processing is patch-based, the deep learning workflow can be distributed in parallel, overcoming the memory problem of large-scale data.
	Our experiments show that the proposed method can effectively suppress few-view artifacts while faithfully preserving textural details.
\end{abstract}

\section{Introduction}

	Sparse-view CT remains an important topic in both industrial and academic contexts, because it offers several unique advantages; for example, it potentially reduces the radiation dose to a patient while maintaining the diagnostic performance. As a result, a good number of researchers studied this problem to suppress image artifacts and noise for clinical applications.
	
	In sparse-view CT reconstruction, due to the angular down-sampling that does not satisfy Nyquist's criterion, the analytical reconstruction algorithm -- filtered back-projection (FBP) cannot obtain satisfactory results.
	Sparse-view CT reconstruction can also be viewed as solving a large linear inverse problem. However, classic iterative algorithms for CT reconstruction, such as algebraic reconstruction technique (ART)~\cite{gordon1970algebraic}, simultaneous algebraic reconstruction technique (SART)~\cite{andersen1984simultaneous}, expectation maximization
	(EM)~\cite{dempster1977maximum}, etc., cannot achieve satisfactory results, given the under-determined and ill-conditioned nature of the problem.
	
	Compressed sensing (CS) theory~\cite{candes2006robust, donoho2006compressed} provides a solid basis for much-improved sparse-view CT reconstruction. CS theory proves that if a signal is sparse and the sampling process satisfies the restricted isometry property (RIP), then the signal can be perfectly recovered with an overwhelming probability. Inspired by CS theory, researchers proposed various sparsity-promoting iterative reconstruction (IR) algorithms for sparse-view CT reconstruction. Hu \textit{et al.} introduced total variation (TV) for the sparse regularization, resulting in encouraging results~\cite{yu2005total}. To overcome the over-smoothness caused by TV regularization, the generic TV method was modified into quite many variants, including anisotropic TV~\cite{chen2013limited}, total generalized variation (TGV)~\cite{niu2014sparse}, fractional order TV~\cite{zhang2014few, zhang2016statistical}, and so on. In addition to the sparsifying transform based on discrete gradients used in TV, other sparsifying transforms were also introduced for sparse-view CT reconstruction. In~\cite{xu2012low, chen2013improving}, a learned redundant dictionary was used for a sparsifying transform.  Similarly, sparsifying transforms were trained on high-quality external CT images to obtain promising results~\cite{zheng2016low, chun2017sparse}. Moreover,
	a tight framework transform was proposed to represent sparse signals for multi-energy CT reconstruction~\cite{gao2011multi}.
	
	The above-mentioned studies demonstrate that CS-based sparse-view CT reconstruction can achieve satisfactory results. However, these methods have significant drawbacks as well. In clinical practice, there are differences in tissue composition at different locations in a patient, and among individuals. It is thus difficult to obtain a universal regularization term to handle diverse images optimally. The reconstruction performance of these IR algorithms also depends on the hyper-parameters, whose selection requires expertise and experience, and needs to be done case-by-case. Furthermore, the computational cost of these algorithms is very high. These problems are road blocks towards clinical translation.
	
	Recently, deep learning (DL)~\cite{lecun2015deep} has transformed many fields including medical imaging. By learning from a large number of data samples, DL extracts intrinsic features from the data. DL was first demonstrated extraordinarily successful in natural language processing (NLP), computer vision, image processing, etc. Inspired by these applications of DL, researchers introduced DL into medical imaging. For example, Chen \textit{et al.} proposed to use a three-layer convolutional neural network (CNN) for low-dose CT denoising~\cite{chen2017lowdose}. Later, Chen \textit{et al.} continued to introduce the residual structure into CNN, and proposed the residual encoder decoder CNN (RED-CNN)~\cite{chen2017low}. Similarly, Jin \textit{et al.} used the U-Net~\cite{ronneberger2015u} to remove sparse-view CT image artifacts~\cite{jin2017deep}. Subsequently, a series of gradually improved U-Nets was proposed for image denoising and artifact removal~\cite{kang2017deep, han2018framing}. To align deep learning-based results with human visual perception, Yang \textit{et al.} incorporated a generative adversarial network (GAN) and a perceptual loss function~\cite{yang2018low}, yielding visually better images. However, these post-processing networks cannot perform satisfactorily in removing sparse-view CT image artifacts. Introducing the constraints of projection data helps improve the capability of the network for few-view CT. Along this direction, Chen \textit{et al.} unrolled the gradient descent algorithm into a deep network and performed the correction of projection data, achieving promising performance~\cite{chen2018learn}. Subsequently,  different IR algorithms were unrolled into networks to improve artifact suppression. For example, the fast iterative shrinkage-thresholding (FISTA) algorithm was unrolled into a network, improving the efficiency of each iteration block so that the use of fewer iteration blocks still led to a good performance~\cite{xiang2021fista}. In ~\cite{wu2021drone, chun2020momentum, wu2017iterative}, networks were designed and trained as priors, and then incorporated into IR algorithm schemes. Moreover, in~\cite{cheng2020learned}, the down-sampling and full-sampling Radon transform were alternately used in a network to recover projection information and improve sparse-view CT reconstruction. These methods all improved the few-view CT image quality, indicating the potential of deep learning in this area. However, these methods often suffer from loss of image details, clearly demanding further improvement. Also, most of these methods work in the supervised learning mode, requiring pairs of inputs and labels that are difficult to obtain clinically. On the other hand, as the main unsupervised learning method for medical imaging, GAN suffers from mode collapse and image hallucinations. Also, these methods use a large amount of memory to store feature maps of images and sinograms, being infeasible in the cases of large-scale images and data.
	
	Over the last two years, the denoising diffusion probabilistic model (DDPM)~\cite{sohl2015deep, ho2020denoising, croitoru2022diffusion, yang2022diffusion} emerged in the image generation field, with remarkable successes and often beyond the GAN performance.
	DDPM gradually adds Gaussian noise to an image, transforms the image through the latent spaces to a Gaussian distribution, and then uses a network to learn the denoising process to trace the latent spaces backward for a generated image out of the original distribution. Impressively, DDPM overcomes the mode collapse problem of GAN and exhibits better stability than GAN in image processing tasks~\cite{rombach2022high, saharia2022image, lugmayr2022repaint, song2021solving}.
	
	Inspired by DDPM, in this paper we propose a projection-domain patch-based DDPM method for sparse-view CT reconstruction. In the training stage, we use a U-Net~\cite{ho2020denoising} to learn the reverse diffusion process for fully sampled projection patches.
	In the sampling stage, we first implement a fully sampling Radon transform on sparse-view CT images to obtain pseudo fully sampled projection data.
	Then, pseudo fully sampled projection data is cropped into patches as the condition for the reverse diffusion process. The patches restored by ordinary differential equation (ODE) sampling are put together to form the final projection dataset, and then high-quality images can be directly reconstructed using filtered back-projection (FBP). The reconstruction with our proposed method can eliminate image artifacts and yet preserve clinically important details. Also, two novel features of our method make it clinically friendly. First, our method does not require paired data, and both training and sampling steps are done in an unsupervised mode. Second, our method is patch-based, which divides a large-scale dataset into essentially independent patches or cubes for parallel computation. These features qualify our method in solving large-scale deep reconstruction tasks such as for high-resolution breast cone-beam CT.
	 
\section{Methodology}

\subsection{Score-Based DDPM}
	We denote a fully sampled projection dataset as $ \bm{Y} \in \mathbb{R}^{N_v\times N_d} $, where $ N_v $ and $ N_d $ represent the number of projection views and the number of detector elements respectively. The down-sampled projection data can be obtained by a linear transform 
	\begin{equation}
		\bm{Z} = P(\bm{M} \odot \bm{Y}),
		\label{eq:1}
	\end{equation}
	where $ \bm{Z} \in \mathbb{R}^{N_v'\times N_d} $ denotes the sub-sampled projection dataset, $ \bm{M} \in \mathbb{R}^{N_v\times N_d} $ is the mask, with the entry $ \bm{M}_{ij} = 1 $ if the $ i $-th view is sampled, otherwise $ \bm{M}_{ij} = 0 $, and $ \odot $ represents the element-wise multiplication, and $ P: \mathbb{R}^{N_v\times N_d} \rightarrow \mathbb{R}^{N_v'\times N_d} $ is the operation that extracts the selected view data from the original projection dataset.
	
	To perform the patch-based diffusion, we randomly extract a patch $ \bm{y} \in \mathbb{R}^{d\times d} $ from the fully sampled projection dataset $ \bm{Y} $. According to~\cite{ho2020denoising}, the forward process of DDPM is a Markov chain which gradually adds Gaussian noise to the clean patch $ \bm{y}_0 = \bm{y}$ with a predefined variance sequence $ \bm{\beta} = \{\beta_1, \beta_2, \cdots, \beta_T \} $:
	\begin{equation}
		q(\bm{y}_{1:T}|\bm{y}_{0}) = \prod_{t=1}^T q(\bm{y}_t|\bm{y}_{t-1}),
		\label{eq:2}
	\end{equation}
	where
	\begin{equation}
		q(\bm{y}_t|\bm{y}_{t-1}) = \mathcal{N} (\bm{y}_t| \sqrt{1 - \beta_t}\bm{y}_{t-1}, \beta_t \bm{I}).
		\label{eq:3}
	\end{equation}

	Due to the properties of the Gaussian distribution, the iteratively perturbed patch at any time step $ t $ can be directly obtained from $ \bm{y}_0 $:
	\begin{equation}
		q(\bm{y}_t|\bm{y}_{0}) = \mathcal{N} (\bm{y}_t| \sqrt{\bar{\alpha}_t}\bm{y}_0, 	(1-\bar{\alpha}_t) \bm{I}),
		\label{eq:4}
	\end{equation}
	where $\alpha_t = 1-\beta_t$ and $ \bar{\alpha}_t = \prod_{i=1}^t \alpha_i $. The final result of the forward process will approach the normal distribution $ \bm{y}_T  \sim \mathcal{N}(0,\bm{I}) $.
	
	Clearly, the specific implementation of Eq. (\ref{eq:4}) is as follows:
	\begin{equation}
		\bm{y}_t = \sqrt{\bar{\alpha}_t}\bm{y}_0 + \sqrt{1-\bar{\alpha}_t} \bm{\epsilon}, \ 	\bm{\epsilon}\sim \mathcal{N}(0, \bm{I}).
		\label{eq:5}
	\end{equation}

	Technically, a U-Net~\cite{ho2020denoising} can be used to learn the Gaussian perturbations involved in the diffusion process:
	\begin{equation}
		\mathcal{L} = \mathbb{E}_{\bm{y}, \bm{\epsilon}, t} \left\| 	\bm{\epsilon} - \bm{\epsilon}_{\bm{\theta}}(\sqrt{\bar{\alpha}_t}\bm{y}_0 + \sqrt{1-\bar{\alpha}_t} \bm{\epsilon}, t) \right\|_2^2.
		\label{eq:6}
	\end{equation}

	In the inference stage, the reverse diffusion process $ \bm{y}_{t-1} \sim p_{\theta}(\bm{y}_{t-1}|\bm{y}_{t})$ can be iteratively computed as follows:
	\begin{equation}
		\bm{y}_{t-1} = 	\frac{1}{\sqrt{\bar{\alpha}_t}}\left(\bm{y}_t-\frac{1-\alpha_t}{\sqrt{1-\bar{\alpha}_t}}\bm{\epsilon}_{\bm{\theta}} (\bm{y}_t, \bm{x}, t)\right) +\sigma_t \bm{\xi}, \ \bm{\xi}\sim \mathcal{N}(0,\bm{I}).
		\label{eq:7}
	\end{equation}
	
	Song \textit{et al.} proposed a method to construct the diffusion process for the continuous time variable $ t\in [0, 1]$, which allows a tractable form for more efficient sampling~\cite{song2020score}. Also, Song \textit{et al.} used the following stochastic differential equation (SDE) to describe the diffusion process~\cite{song2020score}:
	\begin{equation}
		\mathrm{d}\bm{y} = f(t) \bm{y} \mathrm{d}t + g(t)\mathrm{d}\bm{w}, 
		\label{eq:8}
	\end{equation}
	where $ \bm{w} \in \mathbb{R}^{d \times d} $ is the Wiener process, $f: \mathbb{R} \rightarrow \mathbb{R}$ is a scalar function to define the drift component, and $g: \mathbb{R} \rightarrow \mathbb{R}$ is another scalar function as the diffusion coefficient.
	According to ~\cite{ho2020denoising}, the reverse diffusion process can also be modeled as the solution to an SDE:
	\begin{equation}
		\mathrm{d}\bm{y} = \left[f(t) \bm{y} -g(t)^2 \nabla_{\bm{y}}\log{p_t(\bm{y})} \right]\mathrm{d}t + g(t)\mathrm{d}\bar{\bm{w}}, 
		\label{eq:8_1}
	\end{equation}
	where $ \bar{\bm{w}} \in \mathbb{R}^{d \times d} $ is another Wiener process for time-reversed SDE, and $ \nabla_{\bm{y}}\log{p_t(\bm{y})} $ is referred to as the score. Once the score of each marginal distribution $ \nabla_{\bm{y}}\log{p_t(\bm{y})} $ is known, a high-quality patch can be obtained by time-reversed SDE sampling from $ \bm{y}_T  \sim \mathcal{N}(0,\bm{I}) $.
	
	Just like DDPM that learns each incremental noise perturbation, the score-based model uses a network to estimate the score. Hence, we can train a time-dependent network score estimation model $ \bm{s}_{\theta} $ with the following loss function:
	\begin{equation}
		\mathcal{L} = \mathbb{E}_{\bm{y}, t} \lambda (t)\left\| \bm{s}_{\bm{\theta}} (\bm{y}(t),t)	- \nabla_{\bm{y}(t)}\log{p_t(\bm{y}(t)|\bm{y}(0))} \right\|_2^2.
		\label{eq:8_2}
	\end{equation}
	where $ \lambda (t) $ is a positive weighting function, $ \bm{y}(0) \sim p(\bm{y}) $ and $ \bm{y}(t) \sim p_t(\bm{y}(t)|\bm{y}(0)) $. According to ~\cite{song2019generative, song2020score}, $ \lambda (t) \propto 1/ \mathbb{E}[\left\|\nabla_{\bm{y}(t)}\log{p_t(\bm{y}(t)|\bm{y}(0))}\right\|^2_2]$, and $ \lambda (t) $ is chosen as $ \lambda (t) = g(t)^2$.
	
	In the time-reversed SDE sampling, the step size needs to be small to reflect the Wiener process~\cite{platen2010numerical}. Song \textit{et al.} proved that the time-reversed SDE sampling shares the same marginal probability density with an ordinary differential equation (ODE) sampling process~\cite{song2020score}. Song \textit{et al.} called this ODE probability the flow ODE, which is formulated as follows:
	\begin{equation}
		\mathrm{d}\bm{y} = \left[f(t) \bm{y} -\frac{1}{2}g(t)^2 \nabla_{\bm{y}}\log{p_t(\bm{y})} \right]\mathrm{d}t, 
		\label{eq:8_3}
	\end{equation}
	By the ODE sampling, the reverse diffusion process reduces the noise generated by the Wiener process and allows a larger step size to improve the sampling efficiency.
	
\begin{figure*}[tbp]
		\centering
		\includegraphics[width=1.0\textwidth]{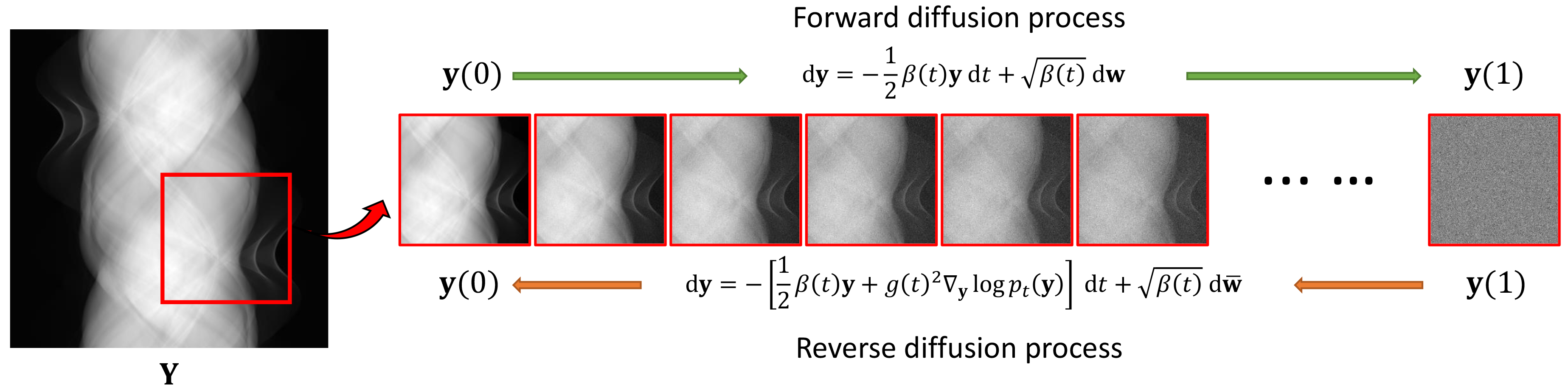}
		\caption{Score-based DDPM for projection patch sampling.}
		\label{fig:1}
	\end{figure*}

	Essentially, DDPM in~\cite{ho2020denoising} can be seen as a special form of SDE.
	In~\cite{song2020score}, Song \textit{et al.} proved that the diffusion process of Eq. (\ref{eq:3}) is essentially equivalent to the following SDE:
	\begin{equation}
		\mathrm{d}\bm{y} = -\frac{1}{2} \beta(t) \bm{y} \mathrm{d}t + \sqrt{\beta(t)}\mathrm{d}\bm{w}, 
		\label{eq:9}
	\end{equation}
	where $\beta(t)$ is the continuous form of the parameter sequence $\bm{\beta}$. The specific expression of $\beta(t)$ for $t\in[0,1]$ is
	\begin{equation}
		\beta(t) = 1000 (\beta_1 + t(\beta_T - \beta_1)).
		\label{eq:10}
	\end{equation}
	By solving Eq. (\ref{eq:9}), the continuous version of Eq. (\ref{eq:4}) can be obtained, which is the following iteratively perturbed patch at any time instant $t$:
	\begin{equation}
		q(\bm{y}(t)|\bm{y}(0)) = \mathcal{N} \left(\bm{y}(t)| a(t)\bm{y}(0), 	b(t)^2 \bm{I}\right),
		\label{eq:11}
	\end{equation}
	where
	\begin{equation}
		a(t) = \exp\left({-\frac{1}{2} \int_0^t \beta(s) \mathrm{d}s}\right), \quad	b(t)^2 = \left(1-\exp\left({- \int_0^t \beta(s) \mathrm{d}s}\right)\right).
		\label{eq:12}
	\end{equation}
Actually, the perturbation prediction network of DDPM $ \bm{\epsilon}_{\theta} $ can regarded as estimating the scaled score $ -\sigma_t \nabla_{\bm{y}}\log{p_t(\bm{y}(t)}$. Therefore, the trained perturbation prediction network and the score prediction network are equivalent. In this paper, we trained a score estimation network on patches extracted from the fully sampled projection dataset. The whole workflow is shown in Fig.~\ref{fig:1}, with Algorithm~\ref{alg:1} describing the training procedure for the estimation model $ \bm{s}_{\bm{\theta}} $.
	
	\begin{algorithm}[t]
		\caption{Training of the score estimation model $ \bm{s}_{\bm{\theta}} $.}
		\label{alg:1}
		\KwIn{Diffusion parameters $\beta_1, \beta_T $}
		\KwOut{Trained model $ \bm{s}_{\bm{\theta}} $}
		\BlankLine
		Initialize $ \bm{s}_{\bm{\theta}} $ randomly;
		
		\While{\textnormal{not converged}}{
			$ \bm{Y} \sim p(\bm{Y}) $
			
			Randomly extract a patch $ \bm{y}(0) $ from $ \bm{Y} $
			
			$ t\sim \mathrm{Uniform}([0,1]) $
			
			Update $\bm{\theta}$ with the gradient $\nabla_{\bm{\theta}} \left[g(t)^2 \left\| \bm{s}_{\bm{\theta}} (\bm{y}(t),t)	- \nabla_{\bm{y}(t)}\log{p_t(\bm{y}(t)|\bm{y}(0))}\right\|_2^2\right]$
		}
	\end{algorithm}

\subsection{Conditional ODE Sampling}
When sampling projection data via time reversal, we sample each data patches in parallel. The process to sample each patch can be formulated as
\begin{equation}
	\bm{y}^{i}(t_{j+1}) = \bm{y}^{i}(t_j) -\left[\frac{1}{2} \beta(t_{j}) \bm{y}^{i}(t_j) +g(t_{j})^2 \bm{s}_{\bm{\theta}} (\bm{y}^{i}(t_j), t_j) \right]\Delta t + \sqrt{\beta(t_j)} \Delta \bar{\bm{w}}, 
	\label{eq:13}
\end{equation}
where $ \bm{y}^{i}(t_j) $ is the $i$-th patch of $ \bm{Y} $ at time $ t_j $ in the time-reversed SDE. The SDE for time reversal can be solved with the Euler-Maruyama sampler and further corrected by the Langevin dynamics, as demonstrated in~\cite{song2020score}. Starting from $ \bm{y}^{i}(t_0) \sim \mathcal{N} (0,\bm{I})$, such a sampling process will eventually generate a random projection patch. To restore a down-sampled projection patch, we will add a real down-sampled projection patch as the condition to the reverse diffusion process. For the down-sampled projection data $ \bm{Z} $, we first use the FBP algorithm to obtain the noisy image $ \bar{\bm{X}} $, then perform the full Radon transform of $ \bar{\bm{X}} $ to obtain a noisy fully sampled projection dataset $ \bar{\bm{Z}} $. The real projection data $ \bm{Z} $ is used to rectify the noisy  fully sampled projection data by inserting the real projection values $ \bm{Z} $ into $ \bar{\bm{Z}} $:
\begin{equation}
	 \tilde{\bm{Z}} = P^{-1}(\bm{Z}) \odot \bm{M} +\bar{\bm{Z}} \odot (1-\bm{M}),
	 \label{eq:14}
\end{equation}
where $ P^{-1}: \mathbb{R}^{N_v'\times N_d} \rightarrow \mathbb{R}^{N_v\times N_d} $ is the operation that reshapes the down-sampled data into the fullly sampled counterpart by inserting zero into the pixels corresponding to the discarded views, and $ \tilde{\bm{Z}} $ is the final pseudo fully sampled projection dataset. Then, with a fixed stride we extract $ N $ overlapped patches from the pseudo full dataset and the full down-sampling mask to obtain two sets of patches $ \{\tilde{\bm{z}}^i\}_{i=1}^{N} $ and $ \{\bm{m}^i\}_{i=1}^{N} $ respectively.

\begin{figure*}[tbp]
	\centering
	\includegraphics[width=0.8\textwidth]{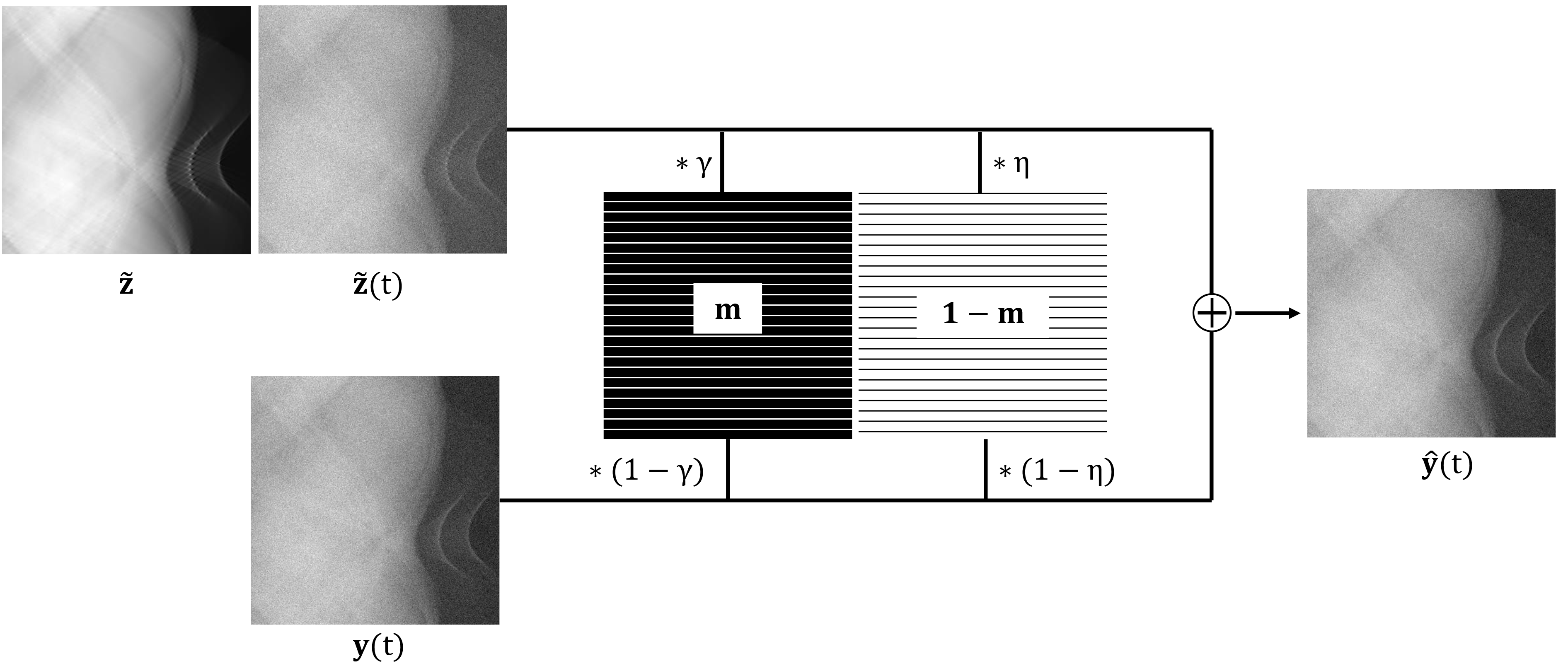}
	\caption{Conditioning method for projection patch sampling.}
	\label{fig:4}
\end{figure*}

For the reverse diffusion process at time $ t_j $, we first obtain the forward diffusion results based on $ \{\tilde{\bm{z}}^i\}_{i=1}^{N} $:
\begin{equation}
	q(\tilde{\bm{z}}^i(t_j)|\tilde{\bm{z}}) = \mathcal{N} \left(\tilde{\bm{z}}^i(t_j)| a(t_j)\tilde{\bm{z}}, 	b(t_j)^2 \bm{I}\right), i=1,2,...,N.
	\label{eq:15}
\end{equation}
These results will be used for the conditional diffusion process. Specifically, inspired by~\cite{song2021solving} we propose a conditioned diffusion method for sparse-view CT projection data restoration. As shown in Fig.~\ref{fig:4}, before implementing Eq. (\ref{eq:13}), the forward diffusion distribution $ \{\tilde{\bm{z}}^i(t_j)\}_{i=1}^{N} $ is used to condition the reverse diffusion sampling as follows:
\begin{equation}
	\hat{\bm{y}}^{i}(t_j) = \left[\gamma \tilde{\bm{z}}^i(t_j) + (1-\gamma)\bm{y}^{i}(t_j)\right] \odot \bm{m}^i + \left[\eta \tilde{\bm{z}}^i(t_j) + (1-\eta)\bm{y}^{i}(t_j)\right] \odot (\mathds{1} -\bm{m}^i),
	\label{eq:16}
\end{equation}
where $ \mathds{1} \in \mathbb{R}^{d\times d} $ is the matrix with all elements being one. The reverse diffusion process conditioned by the pseudo fully sampled projection data is described as Algorithm~\ref{alg:2}.

\begin{algorithm}[t]
	\caption{Inference with the trained denoising model $ \bm{s}_{\bm{\theta}} $.}
	\label{alg:2}
	\KwIn{Number of time steps $T$; diffusion parameters $\beta_1, \beta_T $; under-sampled projection data $ \bm{Z} $; under-sampling mask $ \bm{M} $}
	\KwOut{$\bm{Y}$}
	\BlankLine
	
	Load $ \bm{s}_{\bm{\theta}} $;
	
	$ \bar{\bm{X}} = \mathrm{FBP}(\bm{Z}) $;
	$ \bar{\bm{Z}} = \mathrm{Radon}(\bar{\bm{X}}) $;
	$ \tilde{\bm{Z}} = P^{-1}(\bm{Z})  \odot \bm{M} +\bar{\bm{Z}} \odot (1-\bm{M}) $
	
	Extract patches $\{\tilde{\bm{z}}^i\}_{i=1}^{N} $, $ \{\bm{m}^i\}_{i=1}^{N} $
	
	$ \{t_j\}_{j=1}^T=\mathrm{linspace}(1, 0)$ 
	
	\ForPar{$ i = 1,2,...,N $}{
		$ \bm{y}^i(t_0) \sim \mathcal{N}(0, \bm{I})$
		
		\For{$j=1,2,...,T$}{			
			$ \tilde{\bm{z}}^i(t_j) =  a(t_j) \tilde{\bm{z}} + 	b(t_j) \bm{\xi}, \ \bm{\xi} \sim \mathcal{N}(0,1) $
			
			$\hat{\bm{y}}^{i}(t_j) = \left[\gamma \tilde{\bm{z}}^i(t_j) + (1-\gamma)\bm{y}^{i}(t_j)\right] \odot \bm{m}^i + \left[\eta \tilde{\bm{z}}^i(t_j) + (1-\eta)\bm{y}^{i}(t_j)\right] \odot (\mathds{1} -\bm{m}^i)$
			
			$\bm{y}^{i}(t_{j+1}) = \hat{\bm{y}}^{i}(t_j) -\left[\frac{1}{2} \beta(t_{j}) \hat{\bm{y}}^{i}(t_j) +g(t_{j})^2 \bm{s}_{\bm{\theta}} (\hat{\bm{y}}^{i}(t_j), t_j) \right]\Delta t + \sqrt{\beta(t_j)} \Delta \bar{\bm{w}}$
		}
	}

	Obtain $\bm{Y}$ by assembling patches $\{\bm{y}^{i}(t_T)\}_{i=1}^{N} $ into a full projection dataset.
\end{algorithm}
	
	However, the SDE time reversal perturbs projection data with the Wiener process. Although these may be indistinguishable to naked eyes, they would spread over the entire image domain after FBP reconstruction, resulting in degraded image quality. To avoid the perturbations by the Wiener process, we adopt the ODE sampling method in this study. In~\cite{song2020score}, Song \textit{et al.} used the RK45 ODE solver~\cite{dormand1980family} for ODE sampling. Here we use a more efficient ODE solver to further improve the sampling performance~\cite{lu2022dpm}.
	
	In~\cite{kingma2021variational}, Kingma \textit{et al.} defined the scalar functions in Eq. (\ref{eq:8}) as follows:
	\begin{equation}
		f(t) = \frac{\mathrm{d} \log{\eta_t}}{\mathrm{d}t}, \qquad g^2(t)=\frac{\mathrm{d}\sigma_t^2}{\mathrm{d} t} - 2\frac{\mathrm{d} \log{\eta_t}}{\mathrm{d}t}\sigma_t^2,
		\label{eq:17}
	\end{equation} 
	where $ \eta_t =  \sqrt{\bar{\alpha}_t}$. Substituting these functions into Eq. (\ref{eq:8_3}), the reverse ODE process can be implemented as
	\begin{equation}
		\mathrm{d} \bm{y} = \left[ f(t) \bm{y} + \frac{g^2(t)}{2\sigma_t}  \bm{\epsilon}_{\bm{\theta}}(\bm{y}(t), t) \right] \mathrm{d} t,
		\label{eq:17.1}
	\end{equation}
	which is a semi-linear ODE~\cite{lu2022dpm}, whose solution can be calculated with the variation of constants formula~\cite{atkinson2011numerical} as follows:
	\begin{equation}
		\bm{y}(t) = \exp\left(\int_{s}^{t}f(\tau) \mathrm{d} \tau\right) \bm{y}(s)+ \int_{s}^{t}\left[ \exp\left(\int_{\tau}^{t}f(r) \mathrm{d}r\right) \frac{g^2(\tau)}{2\sigma_t} \bm{\epsilon}_{\bm{\theta}}(\bm{y}({\tau}), \tau) \right]\mathrm{d}\tau.
		\label{eq:18}
	\end{equation}
	Let $ \lambda_t = \log(\eta_t/\sigma_t) $, Eq. (\ref{eq:18}) can be simplified into
	\begin{equation}
		\bm{y}(t) = \frac{\eta_t}{\eta_s} \bm{y}(s) + \eta_t \int_s^t \left[\left(\frac{\mathrm{d}\lambda_{\tau}}{\mathrm{d}\tau}\right)\frac{\sigma_{\tau}}{\eta_{\tau}}\bm{\epsilon}_{\theta}(\bm{y}({\tau}), \tau)\right]\mathrm{d}\tau.
		\label{eq:19}
	\end{equation}
	The predefined $ \lambda_t $ can be obtained with a strictly decreasing function of $ t $, denoted as $ \lambda(t) $, which has a inverse function $ t = t_{\lambda}(\lambda) $. Then, by changing the time variable $t$ into the parameter variable $\lambda$ and denoting $ \hat{\bm{y}}({\lambda}):= \bm{y}({t_{\lambda}(\lambda)})$ and $ \hat{\bm{\epsilon}}_{\bm{\theta}} (\hat{\bm{y}}_{\lambda}, \lambda):= \bm{\epsilon}_{\bm{\theta}}(\bm{y}({t_{\lambda}(\lambda)}), t_{\lambda}(\lambda))$, Eq. (\ref{eq:19}) can be rewritten as
	\begin{equation}
		\bm{y}(t) = \frac{\eta_t}{\eta_s} \bm{y}(s) + \eta_t \int_{\lambda_{s}}^{\lambda{t}} \left[\mathrm{e}^{-\lambda} \hat{\bm{\epsilon}}_{\bm{\theta}} (\hat{\bm{y}}({\lambda}), \lambda)\right] \mathrm{d} \lambda,
		\label{eq:20}
	\end{equation}
	where the integral $ \int \mathrm{e}^{-\lambda} \hat{\bm{\epsilon}}_{\bm{\theta}} \mathrm{d} \lambda$ is called the exponentially weighted integral of $ \hat{\bm{\epsilon}}_{\bm{\theta}} $~\cite{lu2022dpm}. This integral can be numerically calculated by Taylor expansion. According to the order of the Taylor expansion, Lu \textit{et al.}~\cite{lu2022dpm} provided three solvers for the flow ODE, which are Algorithms \ref{alg:3}, \ref{alg:4} and \ref{alg:5} corresponding to the first-order, second-order and third-order Taylor expansions, denoted as \textit{DPM-Solver-1}, \textit{DPM-Solver-2} and \textit{DPM-Solver-3} respectively. As shown in the pseudo codes of these algorithms, $k$ times of computational complexity is needed for functional evaluation in DPM-Solver-$k$ per step. The higher order solver has a faster convergence speed so that it takes fewer steps to achieve satisfactory results~\cite{lu2022dpm}. Therefore, for a limited number of functional evaluations (NFE), the DPM-Solver-3 is recommended. Algorithm~\ref{alg:6} implements the complete workflow with ODE sampling for sparse-view CT reconstruction.
	
	\begin{algorithm}[t]
		\caption{DPM-Solver-1.}
		\label{alg:3}
		\KwIn{Initial value $ \bm{y}(T) $; time steps $\{t_i\}_{i=0}^{M}$; model $\bm{\epsilon}_{\bm{\theta}}$}
		
		\KwOut{$\tilde{\bm{y}}({t_M})$}
		\BlankLine

		$\tilde{\bm{y}}({t_0}) = \bm{y}(T)$
		
		\For{$i=1,2,...,M$}{
			$h_i = \lambda_{t_i}-\lambda_{t_{i-1}}$
			
			$ \tilde{\bm{y}}({t_i}) = \frac{\eta_{t_i}}{\eta_{t_{i-1}}} \tilde{\bm{y}}({t_{i-1}}) - \sigma_{t_i}\left(\mathrm{e}^{h_i}-1\right) \bm{\epsilon}_{\bm{\theta}}(\tilde{\bm{y}}({t_{i-1}}), t_{i-1})$
		}
		
	\end{algorithm}
	
	\begin{algorithm}[t]
		\caption{DPM-Solver-2.}
		\label{alg:4}
		\KwIn{Initial value $ \bm{y}(T) $; time steps $\{t_i\}_{i=0}^{M}$; model $\bm{\epsilon}_{\bm{\theta}}$} 
		\KwOut{$\tilde{\bm{y}}({t_M})$}
		\BlankLine
		
		$\tilde{\bm{y}}({t_0}) = \bm{y}(T)$
		
		\For{$i=1,2,...,M$}{
			$h_i = \lambda_{t_i}-\lambda_{t_{i-1}}$
			
			$s_i = t_{\lambda} (\frac{\lambda_{t_{i-1}}+\lambda_{t_i}}{2})$
			
			$ \bm{u}_{i} = \frac{\eta_{s_i}}{\eta_{t_{i-1}}} \tilde{\bm{y}}({t_{i-1}}) - \sigma_{s_i}\left(\mathrm{e}^\frac{{h_i}}{2}-1\right) \bm{\epsilon}_{\bm{\theta}}(\tilde{\bm{y}}({t_{i-1}}), t_{i-1})$
			
			$ \tilde{\bm{y}}({t_i}) = \frac{\eta_{t_i}}{\eta_{t_{i-1}}} \tilde{\bm{y}}({t_{i-1}}) - \sigma_{t_i}\left(\mathrm{e}^{h_i}-1\right) \bm{\epsilon}_{\bm{\theta}}(\bm{u}_{i}, s_i)$
		}
		
	\end{algorithm}

	\begin{algorithm}[t]
		\caption{DPM-Solver-3.}
		\label{alg:5}
		\KwIn{Initial value $ \bm{y}(T) $; time steps $\{t_i\}_{i=0}^{M}$; model $\bm{\epsilon}_{\bm{\theta}}$} 
		\KwOut{$\tilde{\bm{y}}({t_M})$}
		\BlankLine

		$\tilde{\bm{y}}({t_0}) = \bm{y}(T),\ r_1=\frac{1}{3},\ r_2=\frac{2}{3}$
		
		\For{$i=1,2,...,M$}{
			$h_i = \lambda_{t_i}-\lambda_{t_{i-1}}$
			
			$s_{2i-1} = t_{\lambda} (\lambda_{t_{i-1}}+r_1h_i),\ s_{2i} = t_{\lambda} (\lambda_{t_{i-1}}+r_2h_i)$
			
			$ \bm{u}_{2i-i} = \frac{\eta_{s_{2i-1}}}{\eta_{t_{i-1}}} \tilde{\bm{y}}({t_{i-1}}) - \sigma_{s_{2i-1}}\left(\mathrm{e}^{r_1 h_i}-1\right) \bm{\epsilon}_{\bm{\theta}}(\tilde{\bm{y}}({t_{i-1}}), t_{i-1})$
			
			$ \bm{v}_{2i-1} = \bm{\epsilon}_{\bm{\theta}}(\bm{u}_{2i-1}, s_{2i-1}) -  \bm{\epsilon}_{\bm{\theta}}(\tilde{\bm{y}}({t_{i-1}}), t_{i-1})$
			
			$ \bm{u}_{2i} = \frac{\eta_{s_{2i}}}{\eta_{t_{i-1}}} \tilde{\bm{y}}({t_{i-1}}) - \sigma_{s_{2i}}\left(\mathrm{e}^{r_2 h_i}-1\right) \bm{\epsilon}_{\bm{\theta}}(\tilde{\bm{y}}({t_{i-1}}), t_{i-1}) - \frac{\sigma_{s_{2i}} r_2}{r_1}\left(\frac{\mathrm{e}^{r_2 h_i}-1}{r_2h_i}-1\right) \bm{v}_{2i-1}$
			
			$ \bm{v}_{2i-1} = \bm{\epsilon}_{\bm{\theta}}(\bm{u}_{2i}, s_{2i}) -  \bm{\epsilon}_{\bm{\theta}}(\tilde{\bm{y}}({t_{i-1}}), t_{i-1})$
			
			$ \tilde{\bm{y}}({t_i} )= \frac{\eta_{t_i}}{\eta_{t_{i-1}}} \tilde{\bm{y}}({t_{i-1}}) - \sigma_{t_i}\left(\mathrm{e}^{h_i}-1\right) \bm{\epsilon}_{\bm{\theta}}(\tilde{\bm{y}}({t_{i-1}}), t_{i-1}) - \frac{\sigma_{t_i}}{r_2}\left(\frac{\mathrm{e}^{h_i} - 1}{h_i} - 1\right) \bm{v}_{2i}$
		}
		
	\end{algorithm}

\begin{algorithm}[t]
	\caption{Sparse-view CT reconstruction with ODE sampling.}
	\label{alg:6}
	\KwIn{Number of functional evaluations (NFE) $ J $; diffusion parameters $\beta_1, \beta_T $; under-sampled projection data $ \bm{Z} $; under-sampling mask $ \bm{M} $}
	\KwOut{$\bm{X}$}
	\BlankLine
	
	Load $ \bm{s}_{\bm{\theta}} $;
	
	$ \bar{\bm{X}} = \mathrm{FBP}(\bm{Z}) $;
	$ \bar{\bm{Z}} = \mathrm{Radon}(\bar{\bm{X}}) $;
	$ \tilde{\bm{Z}} = P^{-1}(\bm{Z})  \odot \bm{M} +\bar{\bm{Z}} \odot (1-\bm{M}) $
	
	Extract patches $\{\tilde{\bm{z}}^i\}_{i=1}^{N} $, $ \{\bm{m}^i\}_{i=1}^{N} $
	
	$ \{t_j\}_{j=1}^T=\mathrm{linspace}(1, 0), \quad T = \lfloor J / 3 \rfloor + 1$ 
	
	\ForPar{$ i = 1,2,...,N $}{
		$ \bm{y}^i(t_0) \sim \mathcal{N}(0, \bm{I})$
		
		$ \mathrm{NFE} = J $
		
		\For{$j=1,2,...,T$}{			
			$ \tilde{\bm{z}}^i(t_j) =  a(t_j) \tilde{\bm{z}} + 	b(t_j) \bm{\xi}, \ \bm{\xi} \sim \mathcal{N}(0,1) $
			
			$\hat{\bm{y}}^{i}(t_j) = \left[\gamma \tilde{\bm{z}}^i(t_j) + (1-\gamma)\bm{y}^{i}(t_j)\right] \odot \bm{m}^i + \left[\eta \tilde{\bm{z}}^i(t_j) + (1-\eta)\bm{y}^{i}(t_j)\right] \odot (\mathds{1} -\bm{m}^i)$
			
			$ \bm{\epsilon}_{\bm{\theta}}(\cdot, \cdot) = -\sigma_{t_j} \bm{s}_{\bm{\theta}}(\cdot, \cdot)$
			
			\Switch{$ \mathrm{NFE} $}{
				\uCase{1}{
					$\bm{y}^{i}(t_{j+1}) =$ DPM-Solver-1$[\hat{\bm{y}}^{i}(t_j), \{t_j, t_{j+1}\},\bm{\epsilon}_{\bm{\theta}}(\cdot, \cdot)]$
				}
			\uCase{2}{
					$\bm{y}^{i}(t_{j+1}) =$ DPM-Solver-2$[\hat{\bm{y}}^{i}(t_j), \{t_j, t_{j+1}\},\bm{\epsilon}_{\bm{\theta}}(\cdot, \cdot)]$
			}
			\uCase{3}{
				$\bm{y}^{i}(t_{j+1}) =$ DPM-Solver-2$[\hat{\bm{y}}^{i}(t_j), \{t_j, t_{j+1}\},\bm{\epsilon}_{\bm{\theta}}(\cdot, \cdot)]$
				
				$ \mathrm{NFE}= \mathrm{NFE}-2 $
			}
			\Other{
				$\bm{y}^{i}(t_{j+1}) =$ DPM-Solver-3$[\hat{\bm{y}}^{i}(t_j), \{t_j, t_{j+1}\},\bm{\epsilon}_{\bm{\theta}}(\cdot, \cdot)]$
				
				$ \mathrm{NFE}= \mathrm{NFE}-3 $
			}
		
			}
		}
	}
	
Assembling patches $\{\bm{y}^{i}(t_T)\}_{i=1}^{N} $ into the final full dataset $\bm{Y}$ 
	
	$ \bm{X} = \mathrm{FBP}(\bm{Y}) $
\end{algorithm}

	\begin{figure*}[tbp]
		\centering
		\includegraphics[width=1.0\textwidth]{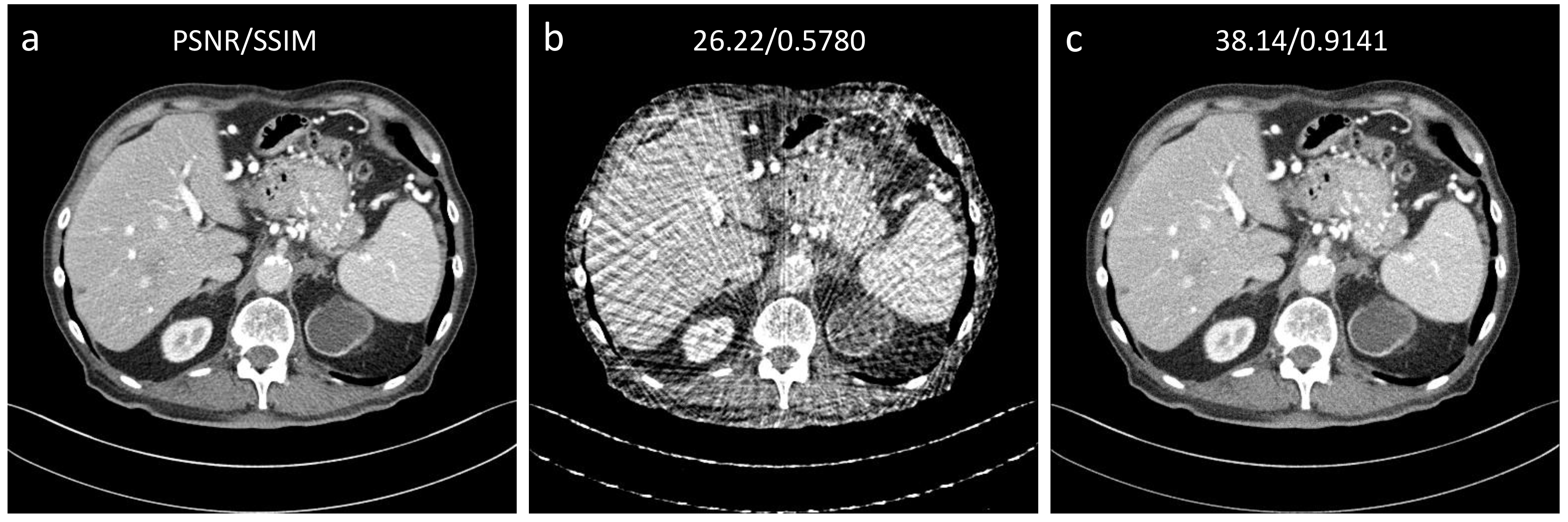}
		\caption{Representative results obtained using our proposed patch-based method. (a) The ground truth, (b) the FBP sparse-view CT reconstruction, and (c) the reconstruction using our patch-based diffusion method. The display window is consistently set to [-160, 240] HU.}
		\label{fig:2}
	\end{figure*}

\section{Experiments and Results}
	To evaluate the performance of our patch-based method for sparse-view CT reconstruction, the \textit{2016 NIH-AAPM-Mayo Clinic Low-Dose CT Grand Challenge} dataset was chosen to conduct initial experiments. The dataset has 2,378 paired CT images with a slice thickness of 3mm from 10 patients. We selected 1,923 paired images from 8 patients as the training set, and 455 paired images from the remaining 2 patients as the test set. The image size is 512x512. Then, simulated projection datasets were obtained with the distance-driven algorithm~\cite{de2004distance}. The distance from the x-ray source focal spot to the isocenter of the imaging field of view is 595 mm.  The distance from the detector to the source is 1085.6 mm. The number of detector elements is 736, each of which has a pitch of 1.2854 mm. The image pixel is 0.6641 mm. In total, 736 projection views were uniformly collected as a full projection dataset. In our method, the patch size is 64x64. For the forward and backward diffusion processes, the continuous time range is $ t\in[0,1] $. $ \beta_1 $ and $ \beta_T $ are  $ 10^{-4} $ and $ 0.02 $ respectively, according to the recommendation in~\cite{ho2020denoising}. The model was trained with the Adam optimizer~\cite{kingma2014adam} at a learning rate of $ 1\times 10^{-4} $. The training process converged well after $ 2\times10^{5} $ iterations on a computing server equipped with an Nvidia RTX A5000 GPU. When sampling, the projection data was uniformly down-sampled to 92 projection views. The stride for extracting patches was set to 32 for overlapped patches. The number of functional evaluations (NFE) was set to 1000. The conditioning parameters $ \gamma $ and $ \eta $ were set to $ 1.0 $ and $ 0.1 $ respectively.
	
\begin{figure*}[tbp]
		\centering
		\includegraphics[width=0.4\textwidth]{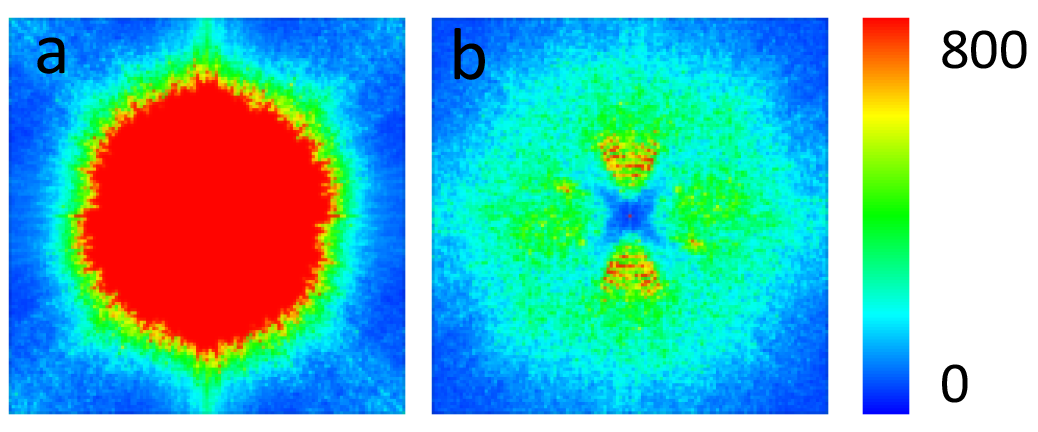}
		\caption{NPS maps of the results in reference to the ground truth. (a) The results for the FBP Sparse-view CT reconstruction, and (b) the counterpart associated with our patch-based diffusion method. The display window is [0, 800] HU$^2$mm$^2$.}
		\label{fig:3}
	\end{figure*}

	Fig.~\ref{fig:2} shows the results obtained with our proposed method. It can be seen that our proposed method eliminated the image artifacts contained in the FBP sparse-view CT reconstruction. In particular, our method also preserves structural features well with neither significant blurring nor false features, which often occur with GAN-based methods. From a perspective of textural analysis, the reconstructed CT image with our proposed method has a texture very similar to the ground truth. To analyze the statistics of reconstruction quality, we calculated the noise power spectrum (NPS) as follows~\cite{tward2008cascaded, dolly2016practical}:
	\begin{equation}
		\mathrm{NPS}(f_i,f_j) = \frac{\Delta_i \Delta_j}{N_i N_j}\langle|\mathrm{DFT}\{\Delta I(i,j)\}|^2\rangle,
		\label{eq:21}
	\end{equation}
	where $\Delta_i=\Delta_j$ is the physical size of the pixel, $N_i=N_j=127$ are the height and width of a region of interest (ROI), and $\Delta I(i,j)$ represents the noise-only realization obtained by subtracting the ground truth from each reconstructed ROI. A total of $129\times129=16641$ ROIs were extracted to calculate NPS. The operator $\langle \cdot \rangle$ denotes the ensemble average over all the ROIs. Fig.~\ref{fig:3} shows the NPS maps in reference to the ground truth. It is seen that the reconstruction error of our method is mainly concentrated in the mid-frequency band, which makes our reconstruction results visually excellent. 

	\begin{table}[t]
		\centering
		\caption{Average PSNR and SSIM results with FBP and our proposed method.}
		\label{tab:1}
		\begin{tabular}{ccc}
			\toprule
			           & PSNR  & SSIM   \\
			\midrule
			Sparse-View CT              & 26.41 & 0.5608 \\
			Ours             & 38.69 & 0.9132 \\
			\bottomrule
		\end{tabular}
	\end{table}

\section{Conclusion}
	In this paper, we have proposed a patch-based DDPM model for sparse-view CT reconstruction. The method has an excellent anti-artifact performance while preserving structural details textural perception. Our proposed method is based on unsupervised learning, overcoming the difficulty in the acquisition of clinical paired data. At the same time, in our method the entire projection dataset is divided into patches so that many patch-based reverse diffusion processes can proceed in parallel, enabling the deep reconstruction in the cases of large-scale datasets such as for high-resolution breast CT and photon-counting CT.

\bibliographystyle{plain} 
\bibliography{reference}

\end{document}